\documentclass[a4paper]{jpconf}
\usepackage{graphicx}
\usepackage{amsmath,amsfonts}
\usepackage[top=4cm,bottom=2.7cm,left=2.5cm,right=2.5cm]{geometry}

\newcommand{\lb}{\langle \kern-.17em \langle} 
\newcommand{\rb}{\rangle \kern-.17em \rangle }

\newcommand{\beq}{\begin{eqnarray}}
\newcommand{\eeq}{\end{eqnarray}}

\begin{document}
\title{Efficient computations of continuous action densities of states for lattice models}

\author{Biagio Lucini$^{1}$, Olmo Francesconi$^2$,
  Markus Holzmann$^3$, David Lancaster$^4$ and Antonio Rago$^{4,5}$}

\address{$^1$ Mathematics Department and Swansea Academy of Advanced
  Computing, Swansea University, Bay Campus, SA1 8EN, Swansea, Wales,
  United Kingdom}
\address{$^2$ IMADA and CP3-Origins, Univ. of Southern Denmark, Campusvej 55, DK-5230 Odense, Denmark}
\address{$^3$ Universit\'e Grenoble Alpes, CNRS, LPMMC, 3800 Grenoble, France}
\address{$^4$ Centre for Mathematical Sciences, University of Plymouth,
  Plymouth, PL4 8AA, United Kingdom}
\address{$^5$ CERN, Theoretical Physics Department, 1211 Geneva 23, Switzerland}

\ead{b.lucini@swansea.ac.uk, francesconi@imada.sdu.dk, markus.holzmann@grenoble.cnrs.fr,
  david.lancaster@plymouth.ac.uk, antonio.rago@plymouth.ac.uk}

\begin{abstract}
The Logarithmic Linear Relaxation (LLR) algorithm is an efficient
method for computing densities of states for systems with a continuous
spectrum. A key feature of this method is exponential error reduction,
which allows us to evaluate the density of states of a system over
hundreds of thousands of orders of magnitude with a fixed level of
relative accuracy. As a consequence of exponential error reduction,
the LLR method provides a robust 
alternative to traditional Monte Carlo calculations in cases in which states
suppressed by the Boltzmann weight play nevertheless a relevant
role, e.g., as transition regions between dominant
configuration sets. After reviewing the algorithm, we will show an
application in U(1) Lattice Gauge Theory  that has enabled us to obtain the most
accurate estimate of the critical coupling with modest computational
resources, defeating exponential tunneling times between metastable
vacua. As a further showcase, we will then present an application of
the LLR method to the decorrelation of 
the topological charge in SU(3) Lattice Gauge Theory near the
continuum limit. Finally, we will review in general applications of
the LLR algorithm to systems affected by a strong sign problem and
discuss the case of the Bose gas at finite chemical potential.   
\end{abstract}

\section{Introduction}
Markov Chain Monte Carlo (MCMC) are particularly well suited in
calculations in which ensemble averages of extensive quantities that
can be expressed explicitly as a 
function of the fields need to be computed. While cases in this
category cover a good cross-section of relevant quantities in
physics, they are by no means exhaustive. In fact, there are various
scenarios, ranging from computations of partition functions to the
sign problem, in which MCMC are inefficient. 
In this contribution, we will review the Logarithmic Linear Relaxation
(LLR) algorithm, which enables us to overcome the limitations of MCMC
methods in the presence of small overlaps between domains of relevant
configurations. The superior performance of this algorithm is provided
by exponential error reduction\footnote{We use the expression {\em
    exponential error reduction} to indicate that the relative error
  of a quantity is independent from the value of the latter.}, which
is a central feature of the method.

This proceeding is organised as follows. We will present the LLR
algorithm for real action systems and provide some example
applications in Sect.~\ref{sect:real}. In Sect.~\ref{sect:complex}, we will
formulate the algorithm for complex action systems and discuss
its application in a numerical study of the Bose gas at finite
chemical potential, which is a popular benchmark model for assessing
the efficiency of an algorithm at circumventing the sign
problem. Sect.~\ref{sect:conclusions} will summarise our contribution and
outline some further directions of investigation.  

\section{The LLR algorithm for real action systems}
\label{sect:real}
Let us start by considering an Euclidean quantum field
theory\footnote{Although for the sake of definiteness in this work we
  use the language of Euclidean quantum field theories, a translation
  of the method to a statistical mechanics context is immediate.} written in a general
form, whose path integral is given by the expression
\begin{equation}
Z(\beta)=\int [ D \phi ] e^{-\beta S [ \phi ]} \ .
\end{equation}
Here, $\phi$ is a field configuration over a lattice $\Lambda$,
$\beta$ is the coupling and $S$ the action. The integral is performed
over all possible values of the fields.  The density of states
associated to a value $E$ of the action $S$ of the model is defined as
\begin{equation}
\rho(E)=\int [ D \phi ] \delta(S[\phi]-E) \ .
\end{equation}
Using this expression for $\rho$, we can reformulate the path integral
as an integral over all possible energy values weighted by the density
of states and the Boltzmann measure:
\begin{equation}
Z(\beta)=\int d E \rho(E) e^{-\beta E}= e^{- \beta F} \ .
\end{equation}
If the density of states is known, free energies and expectation
values are accessible via simple numerical integrations. For instance,
for an observable that depends only on $E$, 
\begin{equation}
 \langle O \rangle=\frac{\int dE \rho(E) O(E) e^{-\beta E}}{\int dE
   \rho(E) e^{-\beta E}} \ . 
\end{equation}
Therefore, an algorithm that enables us to compute $\rho(E)$ will
straightforwardly give us access, through a numerical integration, to
the free energy $F$ and to values of thermodynamic ensemble averages of
observables depending on $E$ as a function of $\beta$. For systems
with discrete energy levels, an algorithm of this type, the celebrated
Wang-Landau algorithm, was provided
in~\cite{Wang:2000fzi}. The LLR method, which was inspired by the latter work,
is an algorithm for the calculation of densities of states in systems
with a continuous
spectrum~\cite{Langfeld:2012ah,Langfeld:2015fua}. This algorithm is
implemented as follows. We start by dividing the 
(continuum) energy interval of the system in $N$ sub-intervals of
amplitude $\delta_E$, with the interval $n$ centered at the value
$E_n$. We then define a piecewise continuous local linear
approximation of $\log \rho(E)$ as  
\beq
\label{eq:piecewise}
\log \tilde{\rho}(E) = a_n \left(E - E_n \right) + c_n
\qquad \mathrm{for~} E_n - \delta_E/2 \le E \le E_n + \delta_E/2 \ ,
\eeq
which is valid for sufficiently small width $\delta_E$ of the energy
sub-intervals. We obtain $a_n$ as the root of the stochastic equation
\begin{equation}
\label{eq:dangle}
  \langle \langle \Delta E \rangle \rangle_{a_n}= \int_{E_n-\frac{\delta_E}{2}}^{E_n+\frac{\delta_E}{2}} \left(E - E_n \right) \rho(E) e^{-a_n E} dE = 0
\end{equation}
using the Robbins-Monro iterative method~\cite{Robbins:1951}
\beq
\label{eq:rm}
\lim_{m \to \infty} a^{(m)}_n  = a_n  \ , \qquad a^{(m+1)}_n=a^{(m)}_n
- \frac{\alpha}{m} \frac{\langle \langle \Delta E \rangle
  \rangle_{a^{(m)}_n}} {\langle \langle \Delta E ^2 \rangle
  \rangle_{a^{(m)}_n}} \ , \qquad \alpha \ \mathrm{constant} \ . 
\eeq
In Eqs.~(\ref{eq:dangle}) and~(\ref{eq:rm}) we have used the
double-angle notation for the expectation of an observable $O(E)$,
which is defined as
\beq
\langle \langle O(E) \rangle \rangle_{a_n}= \frac{1}{\cal N}
\int_{E_n-\frac{\delta_E}{2}}^{E_n+\frac{\delta_E}{2}} O(E) \rho(E)
e^{-a_n E} dE \ , \qquad {\cal N} = \int_{E_n-\frac{\delta_E}{2}}^{E_n+\frac{\delta_E}{2}} \rho(E)
e^{-a_n E} dE \ .  
\eeq
These energy-restricted integrals can be easily reformulated as integrals over the all
spectrum with sharp cut-offs provided by appropriate Heaviside
functions. These sharp cut-offs can be replaced with a smooth Gaussian cut-off~\cite{Pellegrini:2017iuy}. The double-angle expectations are computed with a MCMC
restricted to the relevant energy interval. Note that if we repeat the
Robbins-Monro algorithm starting from different random numbers,
asymptotically, at fixed number of iterations $m$,
the $a_n^{(m)}$ are gaussianly distributed
around the root $a_n = a_n^{(\infty)}$ with a variance
that goes to zero as $m$ increases. This property provides an immediate strategy 
for converting potential systematic errors into statistical errors in
the root finding procedure.

After computing the $a_n$ for all $n$, setting $c_1 = 0$, the
piecewise continuity of $\log \tilde{\rho}(E)$ gives
  \beq
  c_n =  \frac{\delta}{2} a_1+ \delta \sum_{k=2}^{n-1} a_k +
  \frac{\delta}{2} a_n  \ , \qquad n \ge 2 \ .
  \eeq
The numerically determined $\tilde{\rho}(E)$ can be used in path
integral calculations {\em in lieu} of $\rho(E)$. This procedure result
in a systematic approximation error in $\delta_E$ that is however
controlled, since it scales quadratically with the width of the
sub-intervals. Remarkably, our procedure for determining
$\tilde{\rho}$ provides exponential error
suppression~\cite{Langfeld:2015fua}. It is worth noting that the restricted 
sampling discussed above is non-ergodic. Ergodicity can be recovered using the replica
exchange method, as discussed in~\cite{Lucini:2016fid}. 

Exponential error suppression has been proved to be spectacularly
implemented in the method in~\cite{Langfeld:2012ah}, where, using the
piecewise continuous density of states that has been reconstructed
over 250000 orders of magnitude, it has been shown that the
LLR-determined SU(3) lattice gauge theory plaquette agrees with MCMC
calculations over a wide range of $\beta$ values that interpolate
between the strong coupling and the weak coupling regime of the
theory.

\begin{figure}[ht]
\begin{center}
  \includegraphics[width=0.55\textwidth]{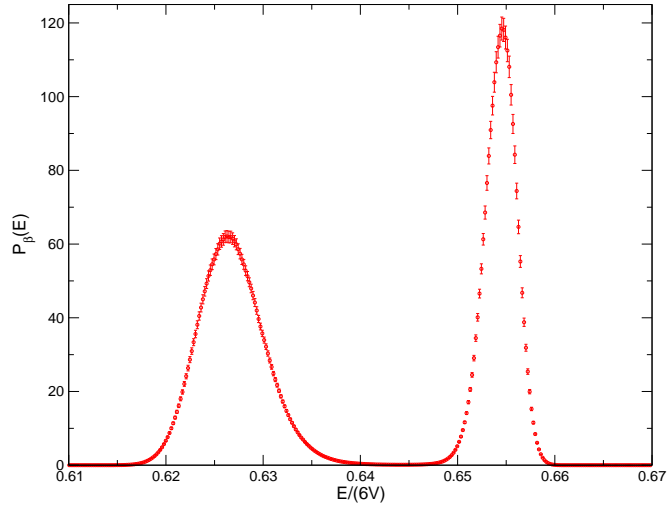} 
\end{center}
\caption{Probability distribution of the energy $E$ for compact U(1)
 lattice gauge theory at $\beta = 1.011006$ (central estimate for the
 critical value using the latent heat peak) on a $20^4$ lattice. \label{fig:2}}
\end{figure}
Due to the presence of a first-order deconfinement phase transition, a
notoriously hard to simulate system is compact U(1) lattice gauge
theory. The best available MCMC calculations used a large amount of
computer time on then state-of-the-art supercomputers, reaching a
maximum lattice size of $18^4$ when periodic boundary conditions were
imposed~\cite{Arnold:2002jk}.  Using the LLR algorithm, we were able
to perform accurate and robust calculations with 
moderate computational resources on a $20^4$ lattice~\cite{Langfeld:2015fua}. A
reconstruction of the probability distribution of the energy at the
pseudocritical value of $\beta$ obtained on our largest system with the LLR algorithm is
shown in Fig.~\ref{fig:2}.

\begin{figure}[ht]
\begin{center}
\includegraphics[width=.7\textwidth]{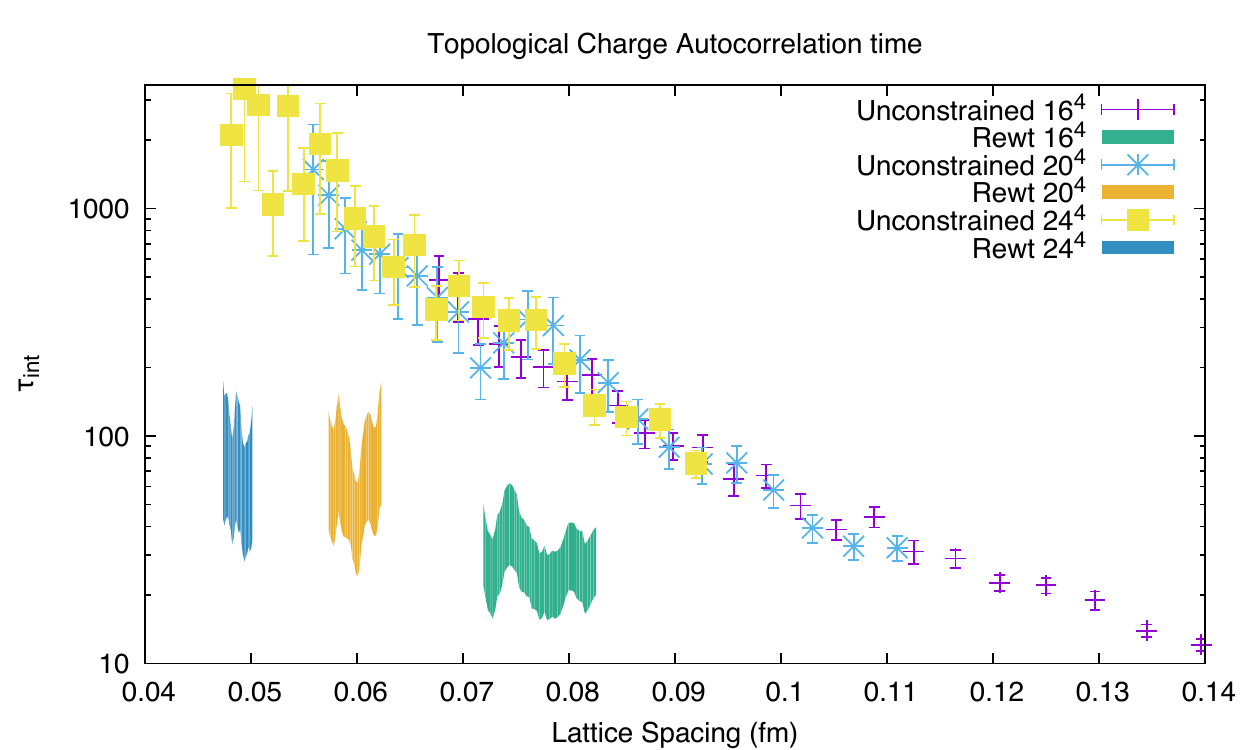}
\end{center}
\caption{Correlation time of the topological charge for SU(3) gauge
  theory at the lattice volumes shown as a function of the lattice
  spacing. The ``unconstrained'' points are obtained with a standard
  heat bath method, while the ``rewt'' points come from LLR simulations.\label{fig:3}}
\end{figure}
More recently, we have shown in~\cite{Cossu:2021bgn} that the LLR
algorithm mitigates significantly
the problem of the topological freezing~\cite{Schaefer:2010hu} by
performing a calculation of the correlation time of the topological
charge in SU(3) pure gauge
theory near the continuum limit. Fig.~\ref{fig:3} shows that, compared
to MCMC methods, the LLR algorithm reduces this correlation time by
about one order of magnitude.

Other successful applications of the LLR method include the calculation of
the order-disorder interface tension in the Potts model~\cite{Lucini:2016fid} and
the calculation of the renormalisation constant for the
energy-momentum tensor in SU(2) lattice gauge theory using the method
of shifted boundary conditions, which in turn requires computations of
free energies from partition functions~\cite{Pellegrini:2017iuy}. 

\section{The LLR algorithm for complex action systems}
\label{sect:complex}
We now consider an Euclidean quantum field theory with a complex action,
whose path integral we write as 
\begin{equation}
Z(\mu)=\int [ D \phi ] e^{-\beta S [ \phi ] + i \mu Q[\phi]} \ .
\end{equation}
The generalised density of states is defined as
\begin{equation}
\rho(q)=\int [ D \phi ]  e^{-\beta S [ \phi ]} \delta(Q[\phi]-q)  \ .
\end{equation}
In terms of $\rho(q)$, we rewrite $Z$ as
\begin{equation}
Z(\mu)=\int d q \rho(q) e^{i\mu  q} \ .
\end{equation}
In general, the above integral is strongly oscillating, These
oscillations generate the
large cancellations typical of a sign problem scenario. Because of the
latter, when performing a calculation $\rho(q)$ in
general needs to be known with an extraordinary degree of accuracy.

The severity of the sign problem is indicated by the {\em vev} of the
phase factor in the phase quenched ensemble:
\begin{equation}
\label{eq:phasefactor}
  \langle e^{i\mu  q} \rangle = \frac{Z(\mu)}{Z(0)} = e^{- V \Delta F }
\to 0 \qquad \mathrm{exponentially~in~} V \ .
\end{equation}
Eq.~(\ref{eq:phasefactor}) shows that the sign problem can be
reformulated as an {\em overlap problem}, as 
the ensemble average of the phase factor depends on the overlap
of relevant configurations for two different partition functions, which is
exponentially suppressed with the volume. Since the LLR method has
proved to be very efficient at computing regions with suppressed
densities, it is natural to explore how it performs in this
case\footnote{A method for calculations of densities of states for
  systems with complex actions that is
  similar in spirit to the LLR algorithm is the FFA method, first proposed
  in~\cite{Giuliani:2016tlu}.}. We note though that, while the
computation of the density is in the scope of the LLR algorithm,
integration is a separate problem. In particular, in the complex case,
it has been shown 
in~\cite{Langfeld:2014nta} that a piecewise approximation of $\rho$ does not
provide sufficient precision for computing relevant observables. In
the same work, using a $Z(3)$ spin model as a prototype system, the
authors showed that a polynomial interpolation of $\rho$ can be
performed that is stable against the order of the interpolation and at
the same time provides the required
accuracy\footnote{See~\cite{Garron:2017fta} for a related proposal for the
  interpolation of the density based on a cumulant expansion.}.

\begin{figure}[t]
\begin{center}
  \includegraphics[width=0.5\textwidth]{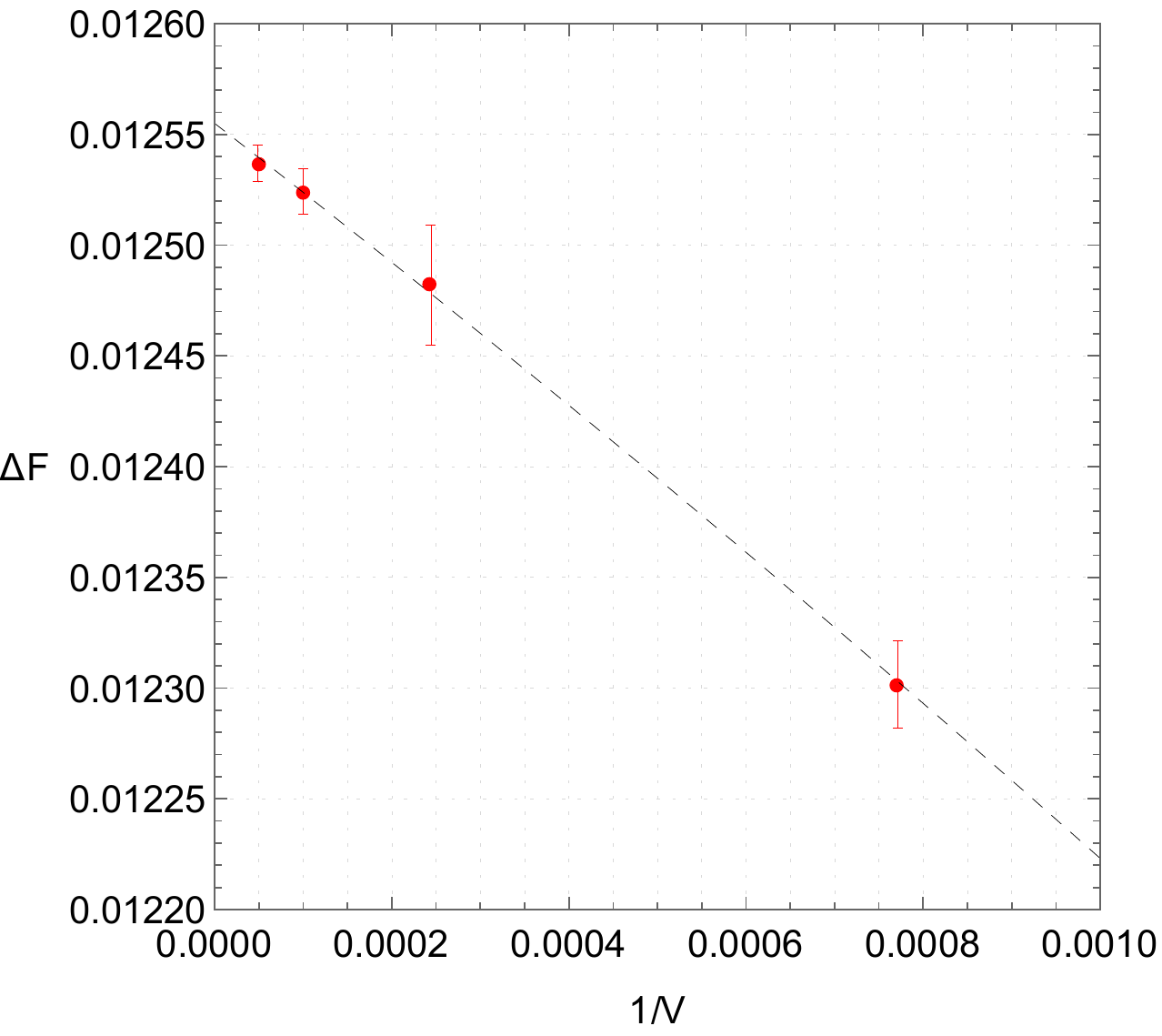}
\caption{Computation of the overlap free energy for the Bose gas at
  finite chemical potential for various volumes. A linear extrapolation
  in $1/V$ to the thermodynamic limit is also shown.\label{fig:4}} 
\end{center}
\end{figure}
More recently, the method has been applied to the four-dimensional
Bose gas system at finite chemical
potential~\cite{Francesconi:2019nph}. Fig.~\ref{fig:4}
shows the computation of the overlap
free energy $\Delta F$ as a function of the volume $V$ for the latter
system at chemical potential $\mu = 0.8$. Volumes up to the size $V =
16^4$ have been used. A linear extrapolation in $1/V$ (also reported in the figure) gives
\begin{equation}
\Delta F = (0.012557 \pm 0.000004) - \frac{(0.329 \pm 0.008)}{V} \ ,
\end{equation}
whose infinite volume limit is reassuringly close to the mean-field
result $\Delta F_{MF}  \simeq 0.012522$ computed
in~\cite{Aarts:2009hn}. At the same time, the LLR result is sufficiently
accurate to expose the expected deviations from the mean-field value. 

\section{Conclusions}
\label{sect:conclusions}
The LLR algorithm enables accurate computations of density of states
that can be used for precise calculations in scenarios in which MCMC
are known to be inefficient. In this contribution, we have reviewed
in particular some applications to first-order phase transitions, to
the decorrelation of the topological charge and to the sign
problem. A natural future direction is an extension to the algorithm
to systems with fermions~\cite{Francesconi:2019aet}. 

\section*{Acknowledgements}
 This work has been partially supported by the ANR project
 ANR-15-IDEX-02e. The work of BL is supported in
part by the Royal Society Wolfson Research Merit Award
WM170010 and by the Leverhulme Foundation Research
Fellowship RF-2020-461\textbackslash 9. BL received funding also from the
European Research Council (ERC) under the European Union’s Horizon
2020 research and innovation programme under grant agreement No 813942.
AR is supported by the STFC
Consolidated Grant ST/P000479/1. Numerical simulations have been
performed on the Swansea SUNBIRD system, provided by the
Supercomputing Wales project, which is part-funded by the European
Regional Development Fund (ERDF) via the Welsh Government, and on the
HPC facilities at the HPCC centre of the University of Plymouth.

\section*{References}
\bibliographystyle{iopart-num}
\bibliography{llr}

\providecommand{\newblock}{}
\begin{thebibliography}{10}
\expandafter\ifx\csname url\endcsname\relax
  \def\url#1{{\tt #1}}\fi
\expandafter\ifx\csname urlprefix\endcsname\relax\def\urlprefix{URL }\fi
\providecommand{\eprint}[2][]{\url{#2}}
% Bibliography created with iopart-num v2.0
% /biblio/bibtex/contrib/iopart-num

\bibitem{Wang:2000fzi}
Wang F and Landau D~P 2001 {\em Phys. Rev. Lett.\/} {\bf 86} 2050
  (\textit{Preprint} \eprint{cond-mat/0011174})

\bibitem{Langfeld:2012ah}
Langfeld K, Lucini B and Rago A 2012 {\em Phys. Rev. Lett.\/} {\bf 109} 111601
  (\textit{Preprint} \eprint{1204.3243})

\bibitem{Langfeld:2015fua}
Langfeld K, Lucini B, Pellegrini R and Rago A 2016 {\em Eur. Phys. J.\/} {\bf
  C76} 306 (\textit{Preprint} \eprint{1509.08391})

\bibitem{Robbins:1951}
Robbins H and Monro S 1951 {\em The Annals of Mathematical Statistics\/} {\bf
  22} 400--407 ISSN 00034851
  \urlprefix\url{http://www.jstor.org/stable/2236626}

\bibitem{Pellegrini:2017iuy}
Pellegrini R, Lucini B, Rago A and Vadacchino D 2017 {\em PoS\/} {\bf
  LATTICE2016} 276

\bibitem{Lucini:2016fid}
Lucini B, Fall W and Langfeld K 2016 {\em PoS\/} {\bf LATTICE2016} 275
  (\textit{Preprint} \eprint{1611.00019})

\bibitem{Arnold:2002jk}
Arnold G, Bunk B, Lippert T and Schilling K 2003 {\em Nucl. Phys. B Proc.
  Suppl.\/} {\bf 119} 864--866 (\textit{Preprint} \eprint{hep-lat/0210010})

\bibitem{Cossu:2021bgn}
Cossu G, Lancaster D, Lucini B, Pellegrini R and Rago A 2021 {\em Eur. Phys. J.
  C\/} {\bf 81} 375 (\textit{Preprint} \eprint{2102.03630})

\bibitem{Schaefer:2010hu}
Schaefer S, Sommer R and Virotta F (ALPHA) 2011 {\em Nucl. Phys. B\/} {\bf 845}
  93--119 (\textit{Preprint} \eprint{1009.5228})

\bibitem{Giuliani:2016tlu}
Giuliani M, Gattringer C and Torek P 2016 {\em Nucl. Phys.\/} {\bf B913}
  627--642 (\textit{Preprint} \eprint{1607.07340})

\bibitem{Langfeld:2014nta}
Langfeld K and Lucini B 2014 {\em Phys. Rev.\/} {\bf D90} 094502
  (\textit{Preprint} \eprint{1404.7187})

\bibitem{Garron:2017fta}
Garron N and Langfeld K 2017 {\em Eur. Phys. J.\/} {\bf C77} 470
  (\textit{Preprint} \eprint{1703.04649})

\bibitem{Francesconi:2019nph}
Francesconi O, Holzmann M, Lucini B and Rago A 2020 {\em Phys. Rev. D\/} {\bf
  101} 014504 (\textit{Preprint} \eprint{1910.11026})

\bibitem{Aarts:2009hn}
Aarts G 2009 {\em JHEP\/} {\bf 05} 052 (\textit{Preprint} \eprint{0902.4686})

\bibitem{Francesconi:2019aet}
Francesconi O, Holzmann M, Lucini B, Rago A and Rantaharju J 2019 {\em PoS\/}
  {\bf LATTICE2019} 200 (\textit{Preprint} \eprint{1912.04190})

\end{thebibliography}

\end{document}